\documentclass[12pt]{article}
\usepackage{graphicx}
\title{Econophysics of a religious cult:\\ the Antoinists  in Belgium [1920-2000] } \vskip2.5cm
\author{\sc  Marcel R.  Ausloos
\\ 
$^1$Beauvallon, r. Belle Jardini\`ere 483, 
B-4031 Li\`ege, Euroland \\
{\it previously at} \\$^2$GRAPES, University of Liege,  
B-4000 Li\`ege, Euroland 
\\ 
\\$^*e$-$mail$ $address$: marcel.ausloos@ulg.ac.be}
\date{}

\begin{document}
\pagestyle{headings}
\maketitle
%\newpage
\pagestyle{plain}
\pagenumbering{arabic}

\begin{abstract}
In the framework of applying econophysics ideas in religious topics,  the finances of  the Antoinist religious movement organized in Belgium between 1920 and 2000 are studied. The interest of investigating financial aspects of such a,  sometimes called, sect  stems in finding characteristics of conditions and mechanisms  under which definitely growth $AND$ decay  features of communities can be understood. The legally reported  yearly income and expenses between 1920 and 2000 are studied. A three wave  asymmetric regime   is observed  over a trend among   marked fluctuations at time of crises. The data  analysis leads to propose a  general mechanistic model taking into account  an average GDP growth, an oscillatory monetary inflation and a logistic population drift. 
\end{abstract}


 \newpage

%\PACS{82.20.Wt, 87.23.Ge,89.75.Hc   ,89.75.Da}
%{Computational modeling; simulation} \and
%{Dynamics of social systems} \and
      %}%{Networks and genealogical trees} \and
   %{Systems obeying scaling laws}

  \begin{figure} 
\centering
\includegraphics[height=10cm,width=14cm]{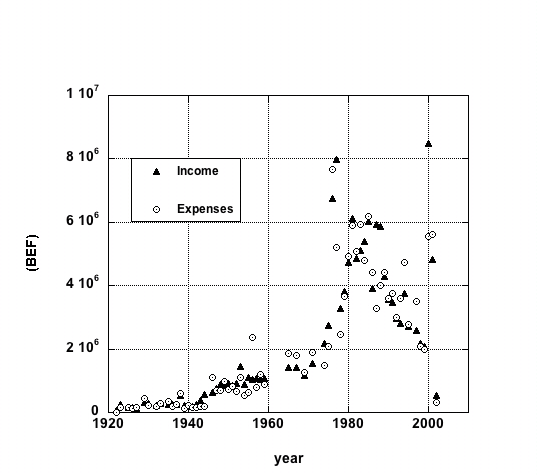}
\caption{Yearly income and expenses of the belgian Antoinist community as reported in the {\it Moniteur Belge} }\label{F1}
\end{figure}

\section{Introduction  }\label{sec:intro} 
 
 In modern statistical physics nowadays \cite{stauffer04},  much work is pertaining to
econophysics \cite{econophysics,carbone,chakr06} and/or sociophysics \cite{chakr06,mimkes,CastellanoFortunatoLoreto}.  Among such topics, religious communities have  attracted some interest  \cite{religion1,religion2,religion566,religion568,garcia,PhA389.10.5479IDPDED_GRMA,mendes}. 
 However the data  is often rare, irregular  and subject to caution. Moreover,
 data acquisition is often relying upon opinion pools or indirect statistical survey means. Their validity  limit  is usually far away  from any accepted one in laboratory studies.\footnote{ For example, the anomalous jump observed  \cite{religion1} at $10^6$ in the number of adherents  for the largest religions, was recently re-pointed out through a  Benford test  by Mir   \cite{TAMir}  on seven major world religions. It is found that the adherent data on all the religions, except Christianity,   conform to the BenfordÕs law. Yet, the significant digit distribution of the three major Christian denominations, Catholicism, Protestantism and Orthodoxy, is found to obey  Benford law.}  Two recent examples can be used for illustrating the point,  i.e. commenting upon (i)  the  evolution of the partial distribution function  or  (ii) the  evolution of  the number of adepts, -  either of  several religious denominations, on the one hand, or  of a given religious movement, on the other hand: 
\begin{itemize}
 \item   
 (i) % the   distribution function of religious denomination evolution can be obtained  \cite{religion1}-\cite{religion568} from 
 the   International Data Base (IDB) \cite{IDB}, 
see table 58, gives  information on the population of 103 nations worldwide, between 1960 and 1992,  specifying the number of adherents for 150 religions. However this is for
 about 2 billions people, i.e. only 1/3 of the present world population; 

(ii)  from the World Christian Encyclopedia \cite{WCT,WCE}, an enormous compilation, one can obtain  the number of adherents  as a function of time, but  only for 56 religions, from 1900 till 2000, -  the data  being averaged over a  regularly distributed 5 years span.

 \item    for a specific  religious movement, i.e. Jehovah's Witnesses,   (i) the  evolution in the number of "publishers", can be examined \cite{StarkIannaccone97jehovah}, but within  5 years averaging windows between 1935 and 1995;
 
 (ii)   the precise and  reliable  high frequency data  used   by Picoli and Mendes \cite{mendes} is for  47 years, i.e. between 1959 and 2005.%to obtain subsequent interpretation of the analysis, due to  the direct implication of one of the authors in the sect.
 \end{itemize}
  
 On the other hand,  from a theoretical point of view, there are plenty of studies about the growth  $or$ the decay of  various ''populations''. Often the approach is based   on $prey$-$predator$ aspects \cite{APCM04}, as   in language  \cite{Patriarca,Schulzereview07}  and more generally "opinion" competitions \cite{dv01,phoenix}. This necessarily  involves two differential equations.  In contrast, the single, concise Verhulst   \cite{Verhulst845}  first order differential equation,   was  proposed to describe the growth  of a community, with a finite life time, and can be adapted to describe a population decay. However such a first order differential equation to describe growth $and$ decay on the same footing seems to be missing \cite{AusloosGV}.  To my knowledge, only Bass model \cite{Bass} describes a growth-death process along   a first order differential equation. It is popular in marketing,  but rarely used, if at all, in  describing the evolution of socio-economic and scientific activities, though apparently governed by similar
growth mechanisms, as often demonstrated.  It is therefore of interest to find examples of  complex systems or processes showing both growth and decay features, and to treat them with in mind a simple first order differential equation as a support. Religions and sects are such organised systems.\footnote{ Interesting growth-decay cases, on a rather short life time,  could be the communist party memberships in various countries, ... but the data on adherents would be highly unreliable, scribit A. P\c{e}kalski}   Thus, the possibility is hereby explored that concepts and methods from statistical
physics may lead to a better understanding of the mechanisms
governing the growth $and$ decay of systems, e.g. of  "churches", ''sects'', and ''religious movements''.
    
    %Moreover, it seems that, in the  scientific literature on the history of religions,  there is no study of a  simple case showing growth and death evolutions.

  In order to do so, it is  necessary to obtain enough reliable data points, over a sufficiently  long time span. However,  religious movements ("organized churches") have a long and often extraordinarily complicated life due to various events, whereas ''sects'' have a very short life and are often somewhat hiding; thus, in both cases,  the data is intrinsically rare and spare, though for different reasons. % Having examined many papers on religions has led me to learn much about the literature on "economics and/of religions"; yet I did not see any data analysis and a quantitative model,  in expert literature, as one that a physicist would like to propose.
In fact, religious movements and sects  
%do not publish data  on adepts for various reasons;  there is
 have very often no binding to publish data on their activities.

 A religious movement community, the Antoinist cult \cite{AntoineL,Vivier,Dericquebourg},  in Belgium, called  a sect in France \cite{JOFrance},   is known to have had a rather huge growth since 1880 or so, but is much decaying nowadays.  Moreover, it is of common knowledge that the evolution in the number of adherents of the Antoinist cult is not due to a prey-predator aspect, but rather stems from socio-economic condition improvements. % In fact, this is in agreement with proposals
 % by Brainbridge and Stark, among others,  have 
%attempting to demonstrate that people do not join and remain in (new) religious movements (sects and cults) on the basis of theological reflection, but only
 % as they are linked to such movements by 
 %because of interpersonal links with group members \cite{LoflandStark1965,Bainbridge1978,StarkBainbridge80}.  %religion is better understood as a social rather than as an individual phenomenon.
  Thus, it can be expected that some interpretation of some  reliable, even $indirect$  $measure$, data might lead to  a description of causes (''forces'') inducing  the overall evolution and  {\it mutatis mutandis}  to suggest ingredients of a model for most, even more complicated,  ''religious'' cases, whatever their size.
 
  The ''exact'' number of adherents to the Antoinist  cult is not known. The more so  since the Antoinist cult \cite{AntoineL,Vivier,Dericquebourg} has  neither proselytism  nor  wealth accumulation as active principles, but rather discretion and anonymous concerns.  This goes against the strong recruitment policy as in all expanding cults, sects and religious movements
\cite{StarkBainbridge80}. Again, this indicates that some non-trivial evolution  of the Antoinist  cult  should be expected \cite{Stark96b}.

% Although the number of adepts can be somewhat estimated, according to police reports or news media, it is  rather rare and highly unreliable, because the cult has no proselytism action, no paid adept, no activity attendance list, and even refuses  to consider such an idea about size counting.  The financial aspects are also minimized.

Fortunately, yearly legal reports about  the budget (expenses and income) of the cult community are known. The data, outlined in Sect. \ref{sec:dataset}, extends between 1922 and 2002, i.e. grossly speaking over the 20-th century. % Since the community after a huge expansion is now decaying,  it is expected that the data could lead to some observation requesting more than a Verhulst  smooth grow-limited approach \cite{Verhulst845}, with some asymptotic value on the population number  after some time.  
   %Thus, in the following, a study is reported about such official data.
     In Sect.\ref{sec:stat}, it is found that $several$ growth-decay regimes  are in fact observable. 
%The  data  presents quite different growth-death magnitudes at different times during the last century.  
 Following an empirical analysis, a concise mechanistic-like model
   %hopefully valid for other sect expansion-decay cases 
  is proposed.  It appears that the data  can be understood   taking into account  various
  %antagonistic (economic and social) 
  factor responses.   A conclusion is given in Sect.\ref{sec:conclusion}.

\section{The data set }\label{sec:dataset}
 The Antoinist cult \cite{AntoineL,Vivier,Dericquebourg}, - see also  $culteantoiniste.com$,  developed during the life time of  ''Father'' (Louis) Antoine (1846-1912), mainly in Belgium and France \cite{JOFrance}.  Father Antoine, after some  catholic education, and some spiritism activity, turned toward predication and ritual  healing. His "philosophy", based on ten "principles" \cite{AntoineL,Vivier,Dericquebourg}, mainly appealed to steel and coal mine workers, first near Li\`ege, Belgium. Then, the religious movement steadily expanded, as can be observed from the evolution in the number of  ''temples'', - approximately 32 in each country\footnote{A few temples are also found in Congo (DRC) and Brazil.} and ''reading rooms'' \cite{Dericquebourg} at the end of the  30's.  
 
The community structure is  legally recognised as an {\it Etablissement d'utilit\'e publique} (Organism of Public Utility) in Belgium,  since 1922. Thus, it must legally report regularly, every year, financial data, approved by the Cult Administration Board and its General Assembly. Such data must be published in the
  Belgian yearly official journal  ({\it Moniteur Belge}), under the supervision of the Minister of Justice,  in charge of  ''churches'' and ''religious movements'' activities in Belgium. 
  
 % In   order to keep the  same monetary unit throughout the analysis, i.e. the Belgian Franc (BEF), only the data till 2000, i.e. before EUR existence,  is  discussed here below. 
  The here below analysed  financial data set  was obtained through the  {\it Moniteur Belge} issues, when available in the archives of the Antoinism Library in Jemeppe-sur-Meuse, Belgium.   There are a few missing few data points. However, they are without much loss  of content for the present discussion and conclusion, as  it will be apparently seen below. Moreover due to a change in regulation, since 2002,  the data has not to be published anymore  in the  {\it Moniteur Belge}, but deposited at some court office in Li\`ege.  In order to  obtain them, one  needs specific approbation from the Antoinist cult hierarchy, approbation which was presently refused. Thus only data up to 2000 are examined below.

 The  income is almost originating in (anonymous) gifts.  The sales of books is  included in such  yearly accounts, but  they  turn out  to be a weak contribution to the total sum, - after examining detailed data, not shown, before 1940.  Almost all incomes are turned toward expenses, which are mainly for the construction and maintenance of the temples. It is important to stress that the so called  ''temple desservants'' and "lecture room readers'' are not paid.   The difference between income and expenses goes into savings accounts, - savings which can be later on used. As seen on Fig.1 sometimes the expenses can turn out  to be larger than the income, but the yearly budget has always been balanced.

  \begin{figure} 
\centering
\includegraphics[height=10cm,width=12cm]{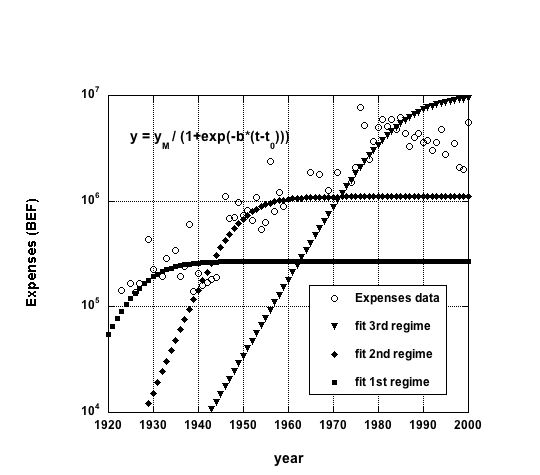}
\caption{Expenses  (logarithmic scale) of  the Antoinist community in Belgium suggesting three growth regimes; they are recognised through fits  with a logistic function on selected time intervals  }
\label{fig:Plot501fitexpenses1stregimebw}
\end{figure}

  \begin{figure} 
\centering
\includegraphics[height=10cm,width=12cm]{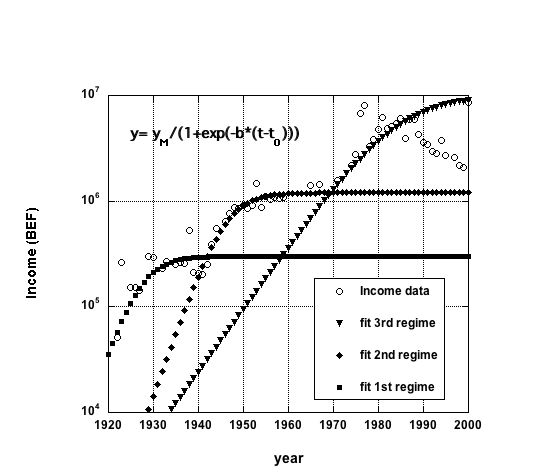}
\caption{Income  (logarithmic scale) of the Antoinist  community in Belgium suggesting three growth regimes; they are recognised through fits  with a logistic function on selected time intervals  }
\label{fig:Plot501fitincome1stregimebw}
\end{figure}

\section{Data    analysis } \label{sec:stat}

\subsection{Data  visual inspection}

The yearly  income and expenses raw data, Fig.\ref{F1},  appear  as pretty scattered. %It seems at once, as mentioned here above to discard  strange data, see Fig.\ref{F1},  given after 2000, when the EUR was introduced. 
  Notice that, yearly offerings can become very large: already a million of BEF in 1960.  Yet, after visual inspection, it appears that, in both income and expenses cases, three growth-decay  regimes can be  imagined, %see Fig. \ref{fig:Plot6data2CDlogbw}, 
over  weakly overlapping time  interval boundaries: [1920-1939], [1939-1967],  and [1967-2004].   It seems that one should emphasize that the time interval  widths, in which the regimes appear,   follow a simple but remarkable progression, i.e.  19, 28 and 37 years, apparently $ (10+ 9*i) years$, where $i$ indicates the order of the  regimes over the 80 years or so  for which the financial data can be  examined.  

For example, the presence of a maximum is seen near 1929  followed by a short decay  till roughly 1940.   A second growth regime is  followed by a small decay at the end of the {\it golden sixties}. The next regime has a sharp rise which leads to a bump with a maximum near 1985 and seems to end at the beginning of the  21-st  century. The latter regime  is better seen than the previous ones because of the $y-$ axis scale.  In this regime, the evolution in the income and expenses is  more stochastic, with two ''special'' years (1976 and 1977) before and two "very special" years after some sort of maximum.  These incomes are attributed to huge and unusual gifts of deceased persons. %Clearly, from 1985  till 2000, the data shows some decay trend, still going on thereafter, .  

Such a  historical 
evolution can be  first qualitatively understood according to the social situation in Belgium during the 20-th century. Indeed, recall that  Father Antoine, among one of his "principles",  argued that sickness and low mood are merely  questions of faith; since there is no evil, it is  thus simply {\it by faith} that  one can be healed: ... faith in oneself and in the Father, who by imposing hands could pass his fluid to those who could believe to get it and be healed. Such a theory was accepted by many workers, - see on ritual healing \cite{ritualhealing04,ritualhealing06}  as a source of religion. It is easily understood that  such a convincing appeal  had to be reduced when the  health care and other social conditions   of workers improved.  Whence, one expects some growth-decay regime(s) in the number of adepts, which might be reflected in the financial aspects of the community.  Several temples are in fact closing nowadays. %However, there is no data taken about how many adherents attend the services in the various temples.   

%Moreover, the offering, as a form of money by the adepts, is strictly anonymous, and only received in cash form.   for such various reasons,  an enormous decay is expected at some time, and surely nowadays.   However the scientific interest of studying such a community  pertains to the fact that it has growth and decay phases over a finite interval of time, still now.

Thus, e.g., from Fig.\ref{fig:Plot501fitexpenses1stregimebw} and Fig.\ref{fig:Plot501fitincome1stregimebw}, it appears in  semi-log plots,  that three regimes can be conceived in both expenses and income evolutions. A test can be made with the growing logistic  function, i.e. $ y /y_M=  [1+ exp(-b\; (t-t_0))]^{-1}$, with parameters appropriate in each regime. On the one hand, one observes that the  growth   always occurs on a longer interval of time than the decay.  Also,    the  income  growth rate  is, at first sight, comparing  Fig.2 and Fig.3, slightly larger than  the expenses growth rate, except in the 3rd regime, - itself occurring after the  golden sixties,  when the  absolute difference between income and expenses is much larger than in the previous regimes,- see  Fig.1.  

One could also expect that there should be, if the community is well managed, some time delay between the incoming money and the expenses.  Time-delay correlations were looked for but no well marked conclusion could be derived, suggesting a  short term, management of the finances of the community. Thus one concludes on an immediate reinvestment of the income, with some yearly savings, the evolution  of these being not discussed further here.  Nevertheless, the  year boundary between the various income regimes  seems to occur  in  slightly different years, (+/-1), that the one for the expenses regimes.

\subsection{Model}

\begin{figure} 
\centering
\includegraphics[height=10cm,width=12cm]{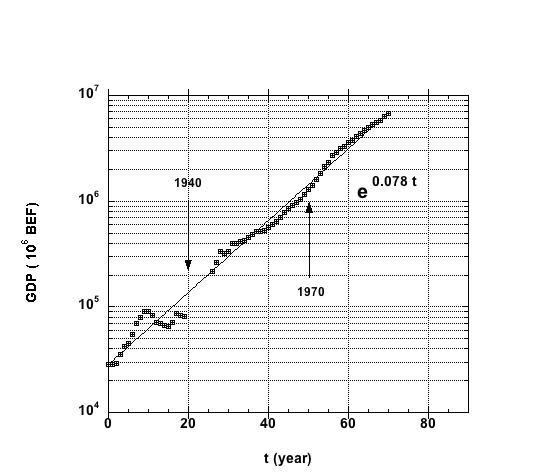}
\caption{Belgium GDP (PPP) suggesting a 3 wave pattern, beside an exponential growth trend; relevant dates for the 20-th century are indicated; time $t_0= 0$ starts in 1920 }
\label{BEGDP}
\end{figure}

It is obvious that the income and expenses span quite different BEF ranges depending on  the time interval. One can note a sort of exponential growth over the century. A search for the trend indicates such an exponential growth indeed. It is therefore reminiscent of the average monetary inflation and GDP growth during the 20-th century:
 \begin{equation}\label{alpha}
  y \simeq    A\;  e^{+\alpha t}   \;\;\;\;\;\; \alpha>0.
\end{equation}
The Belgian GDP growth case is shown in Fig. \ref{BEGDP}, from  estimated \cite{hna}, after being reconstructed  \cite{gd107,gdp107KUL}, 1835-1990 data. It is noticed that $\alpha\simeq 0.078$. On such a figure, some sort of wave, with some fine correspondence to the Antoinist community financial data,  seems also observable, over  the exponential trend.  Thus,  the yearly  incomes and expenses being necessarily positive numbers,  one can  conceive, for  such oscillations, the most simple sort of periodic wave
  \begin{equation}\label{omega}
 y_i \simeq sin(\frac{2\pi t_i}{2 T_i})
\end{equation}
in each  $i$ regime, with $i= 1, 2, 3$, $T_i$ being the corresponding semi-period.

% One effect which might be related  concerns  inflation.  Introduction of such an effect would allow to ''correct'' for the data evolution  in order to extract some specificity of the cult  financial evolution. The relevant data over the time interval of interest. can be found in the Belgium Institute of National Statistics Tables. % \cite{BINS}. Except for the second world war period, during which the data is not available,  and some huge fluctuation  between April 1922 and August 1929,  when it reached - 20\% and +50\% respectively, the inflation in Belgium remained roughly below the 5\% level.  Therefore, even though one has in further analysis and modelling take into account such fluctuations, the present data has not been corrected for such an effect. One should thereby accept that the error bars on the numerical parameter values found here below contain a non stochastic 5\% error.

\begin{figure}[t]
\vskip-2cm
\begin{center}
 \includegraphics[height=10cm,width=14cm]
{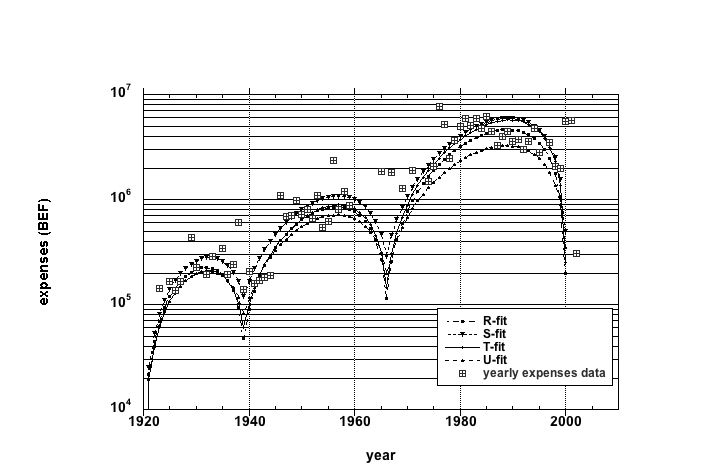}\label{F2} 
\end{center}
\vspace*{8pt}
\caption{Yearly expenses of the belgian Antoinist community as reported in the {\it Moniteur Belge}, with 4 different fits (see text), indicating a three wave structure  }
\end{figure}

\begin{figure}[t]
\vskip-3cm
\begin{center}
 \includegraphics[scale=0.7]
{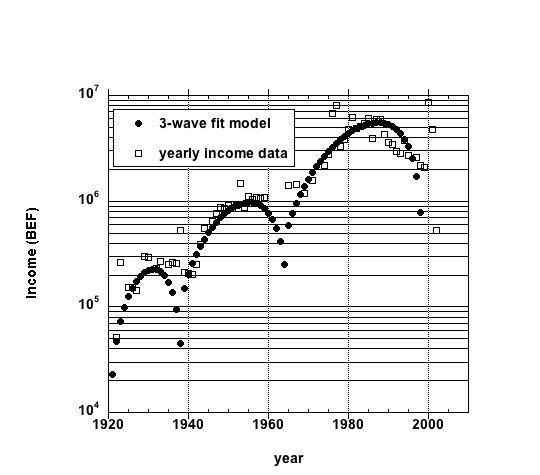}\label{F3}
\end{center}
\vspace*{8pt}
\caption{Yearly income of the belgian Antoinist community as reported in the {\it Moniteur Belge}, with best fit (see text), indicating a three wave structure}
\end{figure}

A second aspect which might be  taken into account %to have a relative view point on the growth and decay phases should be to have some knowledge about 
   pertains to demography,  the possible  change in  the specific population  size of the  areas in which the religious movement was expanding. However this is not available. Some information is known concerning the $whole$ country population;  see e.g. $http://en.wikipedia.org/wiki/$ $Demographics_{-}of_{-}Belgium$. However,  the areas of expansion of the religious movement community were rather limited  from a geographical  point of view, occurring mainly in the industrial areas. However, this possible population effect suggests to introduce a two-parameter logistic law 
   \begin{equation} \label {Verhulst2sol}
      \frac{y}{y_M} =  \frac{1}{1+ e^{-r\; t}}
     \equiv\; \frac{e^{rt}}{1+e^{rt}}\;,
%\label{4}
\end{equation}
 with usual meanings for $y_M$ and $r$,   in order to better  extract some specificity of the religious movement evolution. Moreover, the  introduction of such a Verhulst-like law interestingly allows to reproduce some possible  (though limited)  proselytism effect, due to the so called  ''country capacity'', in a demographic sense,  of the areas.  
 
 Whence, one can introduce a term like
 \begin{equation}\label{beta}
%logistic \;  map 
 \simeq \frac{B_1 e^{\beta t}}{1+ B_2 e^{\beta t}} \sim  \frac{B_3}{1+ B_4 e^{-\beta t}}  \sim (1 + b t)
\end{equation}
 on which the oscillating evolution is superposed.  
 
 In so doing, one may claim to have ingredients depending both on the adept offerings and indirectly their number. They support assumptions in adepts behavior, likely based on imitative behavior  \cite{Bass,Stark96b,hayw011}, and   economic constraints,  for modelling the community finances.

Therefore,  both income and expenses  data are finally proposed to   behave  along  a concise expression 
 \begin{equation}\label{fit}
 y_i  = A_i\;  e^{+\alpha_i t_i}  \; sin\;[\frac{2\pi t_i}{2 T_i}+\phi_i] + (B_i\;  + b_i\; t_i),
\end{equation}
with appropriate parameters, - to be determined in each regime.  Notice that the $\phi_i$ and $B_i$ values can be taken equal to 0, if  the  time origin is chosen at some origin of the time interval  $i$ considered for the fit with the first term 
and  if  a single time origin is always  chosen  for the drift term whatever the time interval ($i=1,2,3$). In practice for the drift term,  the time origin has been imposed to be 1920,  in order to have  simple values of the $x$-axis ticks in the displays.\footnote{One could more exactly choose 1922, but this ''origin of time'' approximation is not likely to be a drastic one, as one can expect.}    Another condition on the fits has been to use integer numbers, both for years and financial  numbers,  these measured in BEF.

Note that  such nonlinear fits lead, as usual,  to  several solutions,  with equivalent  distributions of residuals,  depending on the initial conditions imposed at the start of the trial and error process. Such a situation is illustrated on Fig.5,  for the yearly expenses, where four different fits, with parameters given in Table 1,  over appropriate regime existence intervals, have been  found with rather equivalent  residual distributions, The order of magnitude of the parameters remain $quasi$ the same.  
%Their values are discussed in the conclusion section. 

 The semi-periods $T_i$ of the  sine wave have  been imposed  to be the same one in order to compare each fit meaningfully in each regime, but  yet depending on the regime. Notice that the sum of $T_i$'s for the expense regimes, 38+54+68=160, i.e. correspond exactly  with  the time interval investigated (2000-1920=160/2). Notice also, from the graphs,  a computational detail: in order to minimise the residual spread, the period $T_i$ does not correspond always exactly to half the time interval which is investigated, but  can differ by  a few  years. This might be due to having imposed that $T_i$ be an integer.  In the four    evolution laws for the expenses, the drift  term and the exponential growth  term can be slightly different from each other, however  without much significant difference.

  \begin{table}\label{fitexpensesparameters}\begin{center} \begin{tabular}{|c|c|c|c|c|c|c| c| l |    }
\hline  expenses&  & & &   &  &   \\
\hline\hline   fits &  [$t_0;t_1$]&$A$&$\alpha$&$T$&B &$b$     \\
\hline  R   &[1920;2000] &  $0.11\;10^6$   &0.056&  $-$&0&$0.25\;10^4$   \\
\hline  S   &  $[1920;2000] $ &$0.11\;10^6$    &0.059&   $-$&$0$  &$0.625\;10^4$   \\
\hline  T &$[1920;2000] $&$0.09\;10^6$  &0.062&$-$&0&$1.00\;10^{3.5}$  \\ 
\hline  U &$[1920;2000] $&$0.09\;10^6$  &0.053&$-$&0&$1.375\;10^{3.5}$   \\ 
\hline \end{tabular}  \end{center}
\caption{ Parameter values of  the overall  best fit,  i.e. over the whole data range, see  Fig.5, according to the model equation, Eq.(\ref{fit}), see text,  for  the reported yearly  expenses of the belgian Antoinist community,  assuming 3 different regimes, having each a different  $T_i $, i.e., 38, 54 and 68 years respectively}
\end{table}

   \begin{table}\label{fitincomeparameters}\begin{center} \begin{tabular}{|c|c|c|c|c|c|c| c| l |    }
\hline   income  &  & & &   &  &  \\
\hline\hline   regimes  &  & & &   & &    \\
\hline   $i$&  [$t_0;t_1$]&$A_i$&$\alpha_i$&$2T_i[y]$&$B_i$ &$b_i$      \\
\hline $1 $ &[1922;1940] &  $0.11\;10^6$   &0.059&  $38$&0&$0.265\;10^4$   \\
\hline $2$  &  $[1941;1965]$ &$0.11\;10^6$    &0.059&   $56$&$0$  &$0.615\;10^4$   \\
\hline $3$&$[1966;2000]$&$0.09\;10^6$  &0.059&$74$&0&$1.0725\;10^4$   \\ 
\hline \end{tabular}  \end{center}
\caption{ Parameter values of  the best fits, see  Fig.6,  according to the model equation, Eq.(\ref{fit}), see text,  for  the reported yearly income  of the belgian Antoinist community, assuming 3 different   regimes}
\end{table}
 
 The income case is discussed from Fig.6. A fit following the model equation, Eq.(\ref{fit}),  has been made in three regimes, such that the time intervals  do not overlap, - whence leading to a different $b_i$ parameter. The exponent $\alpha_i$ is found to be the same, within  standard error bars, up to the second decimal. The amplitude $A_i$ is also almost constant for the three regimes, see Table 2.  The periods are slightly longer than for the expenses case, as if for a best fit one should consider  a larger life span [1920-2004].  Still, the overall fit is rather appealing in spite of a  slight anomaly after 1990, and the "very special" [2000-2002] years.
 
 Observe that the growth-decay curves appear asymmetric.\footnote{Such asymmetry between growth and decay also characterises economic cycles \cite{sanglierauslooscycles}.} It is much due to the drift term component of course.
 
 The data  analysis following the suggested model of financial evolution of the community, based on three effects from likely three different causes, leads to reasonable parameters: one fonds (i)    coherence between the time intervals, (ii) the  drift  terms are characterised by a coefficient  $b_i$ leading to a value similar to what is expected in demography, $ \simeq \;0.025$,  concerning the global birth-death rate for the evolution of a population, like  in Belgium during  most of the 20-th century,  and (iii)  the exponent $\alpha$ is  quite similar to that found for the  average evolution of the GDP, e.g. of Belgium,  during the 20-th century, Fig.4.  Note that the   T-fit in Fig. 5 is the best of all four:  in fact it has an $\alpha$ exponent value, the largest  $\simeq 0.062$, and the closest to the one found in Fig. 4 for the Belgium GDP, i.e.   $\simeq 0.078$.

\section{Conclusion } \label{sec:conclusion}

In summary, the objectives of 
the study  were  to find  (i) a  community for which  regular and trusted data exist, using  such financial data to   compensate the lack of knowledge on
the number of adherents,   (ii)  expectedly behaving with a growth and decay history, (iii) such that one can go beyond a Verhulst population description,  but without using a Lotka-Volterra (prey-predator) approach,   (iv)  finding  qualitatively reasonable causes forcing the behaviour in order to describe the evolution along quantitative modelling aspects, and (v) hopefully suggesting a non controversial benchmark case, before studies on more complex community systems.  .

 The income and expenses of the Antoinist cult community in Belgium have been found to fullfil  such objectives, and much more. In the main text, a  good explanation why the 
experimental curves  have such a theoretical form seems to have beeen reached, with very reasonable values of the parameters. 
The ''model'' indicates that  such  religious communities are markedly influenced by external considerations (''external fields''), besides  their intrinsic  ''religious'' goals.  Practically, in the present case, as illustrated, the crash of 1929 induces a drop in income, but the second world war increases the community strength. The golden sixties "reduce" the income:   the adepts well being increased, but the adherents  reduced their offering, becoming in some sense more egoistical. Therefore, one can deduce that  there are two different causes for the drop in income: either a lack of money of the adepts, or in contrast, paradoxically,  ''too much'' wealth. Similarly,  the increase in the religious movement income, at its legal beginning,   may result  from the enthusiastic  thanking for healing the suffering, both of the soul and the body, -   but also  occurs due to the income explosion  until 1985. The variation in expenses are immediately related with such income considerations.

Therefore such a ''model'', - still needing an interpretation of  the so called non-universal  ''amplitudes'',  contains apparently rather ''universal''-like laws, with power exponents,  and is expected  to  be applicable to other societies, - not only religious ones.   It seems 
plausible, that many more societies will be found to have  the same behavior 
and for similar reasons.

In conclusion, it is of interest to study the growth and decay of religious movements, churches, cults from an anthropology \cite{hayw011,StarkBainbridge79,Warren96,hashemi,hayw99,hayw05}, see also \cite{LoganDye84},  or historian \cite{roach},   but also from a sociology \cite{Hamilton01}  or economist \cite{ianna98}   point of view, but combining these also  in studies having   a socio-physics one  \cite{MontrollBadger74}.  
   It  has to be emphasized that the Antoinist cult was appealing because of the suffering of people, working under very hard conditions in the Li\`ege area, when P\`ere Antoine  started to preach and  to give psychological remedies, ''principles'',  for accepting one's life, and demonstrated his healing power.   One crucial aspect of the religious movement concerns its survival under much improved economic, social, and health conditions of workers to whom the P\`ere Antoine's philosophy appealed. Indeed a marked decay in income occurs at the end of the 20-th century, inciting to conclude to a doomed situation, according to the theory in   \cite{Stark96b}, - in contrast to the JehovahÕs Witnesses  \cite{StarkIannaccone97jehovah}. 
 Indeed, ideologies (whether religious or secular) seem to lack coherence and
potency unless they are developed and promulgated by vigorous formal
organisations and social movements  \cite{BainbridgeStark1981}.
Due to the present economic and financial crisis, a phoenix effect \cite{phoenix,futures} might nevertheless take place again.
  
 \bigskip

{\bf Acknowledgements}

 Great thanks to the COST Action MP0801 for financial support. Infinite thanks go to Fr\`ere  W. Dessers  and Soeur S.  Taxquet, 
 %Sylvia Taxquet,  Quai des Ardennes 71 4020 Lige. Tel, 043658896. staxquet@gmail.com
 respectively President and Secretary of the Administration Board of the Antoinist Cult in Belgium, for their kindness, patience,  availability, when I asked for data and historical points, and for  trusting me, -  by allowing me to remove archives from the library for scanning outside the administration  office. Comments by A. P\c{e}kalski have surely  improved this manuscript.


\begin{thebibliography}{99}



 \bibitem{stauffer04} D. Stauffer, 
 Introduction to statistical physics outside physics,   Physica A  336 (2004)  1-5.

 \bibitem{econophysics}   Empirical sciences of financial fluctuations. The advent of econophysics, Tokyo, Japan, Nov. 15-17, 2000 Proceedings H. Takayasu, Ed. (Springer Verlag, Berlin, 2002)% Practical Fruits of Econophysics}, H. Takayasu, Ed., (Springer, Tokyo,  2006).
 
 \bibitem{carbone} A. Carbone, G. Kaniadakis,   A. M. Scarfone, Tails and Ties  Topical Issue on Physics in Society,    Eur. Phys. J. B  57   (2007) 121-125 
 
\bibitem{chakr06} B. K. Chakrabarti, A. Chakraborti, A. Chatterjee.  
Econophysics and Sociophysics: Trends and perspectives. (Wiley-VCH Verlag, Weinheim, 2006). 

\bibitem{mimkes} J. Mimkes,  A Thermodynamic Formulation of Social Science 
in Econophysics and Sociophysics,  B. K. Chakrabarti, A. Chakraborti, and A.  Chatterjee, Eds. (Wiley-VCH, Berlin, 2006).
pp. 279-310.

\bibitem{CastellanoFortunatoLoreto} C. Castellano, S. Fortunato, V. Loreto, Statistical physics of social dynamics, 
Rev. Mod. Phys. 81 (2009)  591Ð646.

\bibitem{religion1} M. Ausloos, F. Petroni,   %Statistical dynamics of religions and  adherents,  
  Europhys. Lett.  77  (2007) 38002.
%  
\bibitem{religion2} M. Ausloos,  F. Petroni,  % Statistical Dynamics of Religions, 
 Physica A   388  (2009)  4438--4444. 
%    
\bibitem{religion566} M. Ausloos, F. Petroni, 
%On World Religion Adherence Distribution  Evolution, 
 in M. Takayasu, T.  Watanabe, H.  Takayasu  (eds.)  
Econophysics Approaches to Large-Scale Business Data and Financial Crisis (Springer, Berlin, 2010) pp. 289-312.  
 
\bibitem{religion568} M. Ausloos,   % On religion and language evolutions seen through  mathematical and agent based models, 
in   C. Rangacharyulu,   E. Haven,  (Eds.)   Proceedings of the First Interdisciplinary CHESS 
Interactions Conference  (World Scientific,  Singapore, 2010)  pp. 157-182 

\bibitem{garcia} A. Garcia Cant\`u Ross, M. Ausloos,   Scientometrics    80  (2009) 457-472.
 
\bibitem{PhA389.10.5479IDPDED_GRMA} G. Rotundo, M. Ausloos, Organization of networks with tagged nodes and
biased links: A priori distinct communities: The case of intelligent design proponents and Darwinian evolution defenders, Physica A 389 (2010) 5479-5494.

 \bibitem{mendes} S. Picoli, Jr.,  R. S. Mendes,
  Universal features in the growth dynamics of religious activities,
     Phys. Rev. E  77 (2008) 036105.

\bibitem{TAMir} T.A. Mir, 
The law of the leading digits and the world religions,
  %Law of the leading digits and the ideological struggle for numbers, 
 Physica A 391 (2012)   792-798.
  
\bibitem{IDB} The International Data Base (IDB) is a computerised
source of demographic and socioeconomic statistics for 227 countries and
areas of the world. The IDB provides a quick access to specialised
information, e.g. ethnicity, religion, and language,  for individual
countries or selected groups of countries in the world. %The major types of data available in the IDB include: Population by age and sex, Vital rates, infant mortality, and life tables, Fertility and child survivorship, Migration,    Marital status, Family planning,       
%Ethnicity, religion, and language, 
%Literacy, Labor force, employment, and income, Households. Sources of the data include:   U.S.Census Bureau, Estimates and Projections,       National Statistics Offices, United Nations and Specialized Agencies (ILO, UNESCO, WHO))

\bibitem{WCT} D. Barrett,  T. Johnson,    World Christian Trends AD 30-AD 2200. Interpreting the annual Christian megacensus  (William Carey Library., Pasadena, 2001).

\bibitem{WCE} D. Barrett, G. Kurian,  T. Johnson,   World Christian Encyclopedia (2nd edition) (Oxford University Press, New York, 2001).

\bibitem{StarkIannaccone97jehovah}  R. Stark,  L.R.  Iannaccone,   Why the JehovahÕs Witnesses Grow so Rapidly: A
Theoretical Application,  J.  Contemp. Relig.  12 (1997) 133-157. 

   \bibitem{APCM04}  A.  P\c{e}kalski,
   A short guide to predator-prey lattice models,
 Comput. Sci.  Eng.  6 (2004) 62-66.

\bibitem{Patriarca} M. Patriarca,   T. Leppanen,
 Modeling language competition,
 Physica A   338 (2004) 296-299.

 \bibitem{Schulzereview07}  %C. Schulze, D. Stauffer, S. Wichmann, Birth, survival and death of languages by Monte Carlo simulation,    Comm. Comput. Phys., nnnnnnnnnnnn
 D. Stauffer, C. Schulze, Microscopic and macroscopic simulation of competition between languages,
 Phys. Life Rev.  2 (2005) 89-116.
 
\bibitem{dv01}
Z. I.  Dimitrova, N.K.  Vitanov,  
 Dynamical consequences of adaptation of the growth  rates in a system of competing populations.
  J. Phys. A: Math. Gen. 34  (2001)  7459 -- 7473.
 
  \bibitem{phoenix} N. K. Vitanov, Z.I. Dimitrova,   M. Ausloos, Verhulst - Lotka - Volterra (VLV) model of ideological struggle, Physica A 389  (2010) 4970-4980.

 \bibitem{Verhulst845}
P.F. Verhulst,
   Recherches math\'ematiques sur la loi d'accroissement de la population.    
  Nouveaux M\'emoires de l'Acad\'emie Royale des Sciences et Belles-Lettres  de Bruxelles 18 (1845)  pp. 1-38. 
 
  \bibitem{AusloosGV}  M. Ausloos, Gompertz and Verhulst frameworks for growth AND decay description, in  Computing Anticipatory Systems, D. M. Dubois, Ed., AIP Conf Proc.  (AIP, Woodbury, 2012)  in press.; $arXiv$ $1109_{-}1269. $
  
  \bibitem{Bass}  F.M. Bass,  
  A new product growth model for consumer durables. innovative and imitative behavior.
  Management Sci.  18  (1969)   215--227.
 
  \bibitem{AntoineL} %L. Antoine, L'enseignement du p\`ere c'est l'enseignement du Christ r\' ev\' el\' e \`a cette \' epoque par la foi  (Presses du Culte Antoiniste, Jemeppe sur Meuse, 0000)
   F. Deregnaucourt, M. Desart, P\`ere Antoine,  R\'ev\'elation par le P\`ere Antoine. Suivi de "Le Couronnement de l'Oeuvre R\'ev\'el\'ee''   (Presses du Culte Antoiniste, Jemeppe sur Meuse, 1909)
 
 \bibitem{Vivier} R. Vivier, D\' elivrez-nous du mal  (Grasset, Paris,  1936)

\bibitem{Dericquebourg} R. Dericquebourg, Les Antoinistes (Brepols, Maredsous,1993)

 \bibitem{JOFrance} Commissions d'enqu\^etes parlementaires sur les sectes en France; Rapport 2468, Dec. 95
 %Bulletin Officiel n¡2000-43 - BULLETIN DU MINISTERE DE LA ...
%antisectes.net/bo-sante.htm
%DƒCRET ET ARRæTƒS RELATIFS Ë LA MISSION INTERMINISTƒRIELLE DE LUTTE CONTRE LES SECTES. Rappel : JO numŽro 234 du 9 octobre 1998, page 15286. 

%French report, 1995, English translation, National Assembly of France, Parliamentary Commission report.
  
      \bibitem{StarkBainbridge80}   R. Stark, W. S. Bainbridge,  Networks of Faith: Interpersonal Bonds and Recruitment to Cults and Sects,  Am. J. Sociol.  85 (1980) 1376-1395.
      
\bibitem{Stark96b} R. Stark,  Why Religious Movements Succeed or Fail: A Revised General Model, J.  Contemp. Relig.  11 (1996) 133-146.

 \bibitem{ritualhealing04} J.  McClenon, The Ritual Healing Theory: Hypotheses for Psychical Research, in  Parapsychology in the twenty-first century: essays  on the future of psychical research,   M. A. Thalbourne and L. Storm, Eds.  (McFarland \& Co., Jefferson, N.C., 2005) pp. 337-360

\bibitem{ritualhealing06} J.  McClenon, The Ritual Healing Theory: Therapeutic Suggestion and the Origin of Religion, in  Where God and science meet: Evolution, genes, and the Religious Brain,  vol.1, How Brain and Evolutionary Studies alter our Understanding    of Religion, 
 P. H. McNamara, Ed. (Praeger Publ., Westport, CT, 2006)  pp.135-158
 
 %McClenon, James. 2006. ÒThe Ritual Healing Theory: Therapeutic Suggestion and the Origin of Religion.Ó In Where God and Science Meet, vol. 1, ed. Patrick McNamara, 135-158. Westport, Conn.: Praeger.
 
\bibitem{hna}  J.P. Smits, P.J. Woltjer,  D. Ma,  A Dataset on Comparative Historical National Accounts, ca. 1870-1950: A Time-Series Perspective, Groningen Growth and Development Centre Research Memorandum GD-107, (University of  Groningen, Groningen, 2009)

 \bibitem{gd107} %Source: For an overview of the sources and research methodology on which the Belgian project is based, see:
  E. Buyst, J.P. Smits,   J.L van Zanden,  National Accounts for the Low Countries: The Netherlands and Belgium, 1800-1990, Scand. Econ. Hist. Rev. 43 (1995) 53 -76.
 
 \bibitem{gdp107KUL}  %For now provisional historical GDP series for Belgium are derived from: 
  E. Horlings, Estimates of Value Added in the Belgian Service Sector 1830-1953 (Leuven: Research Memorandum Catholic University of Leuven, 1996).
  
\bibitem{hayw011}  J. Hayward, Church Growth via Enthusiasts and Renewal, Communication  at the 28th  International System Dynamics Conference, Seoul, South Korea, 2010

 \bibitem{sanglierauslooscycles} M. Ausloos, J. Miskiewicz,  M. Sanglier,
 The durations of recession and prosperity: Does their distribution follow a power or an exponential law?,
  Physica A  339  (2004) 548-558.

% \bibitem{LoflandStark1965} Lofland, J.  and  Stark, R., Becoming a World-Saver: A Theory of Conversion to a Deviant Perspective. nAmerican Sociological Review 30 (1965) 862-875.
 
%  \bibitem{Bainbridge1978}   Bainbridge, W. S.,  SatanÕs Power (University of California Press, Berkeley, 1978)
 
\bibitem{StarkBainbridge79}   J.R. Stark,  W.S. Bainbridge,
    Of Churches, Sects,  and Cults: Preliminary Concepts for a Theory of Religious Movements,  J. Sci. Stud. Relig.  18 (1979) 117-133.      

\bibitem{Warren96} R. Warren, The Purpose Driven Church  (Zondervan Publishing House, Grand Rapids, 1996). 

\bibitem{hashemi} %Fariba 
F. Hashemi,  
An evolutionary model of the size distribution of sect/congregations,
J. Evol. Econ. 10 (2000) 507-521.

  \bibitem{hayw99} J. Hayward,    Mathematical Modeling of Church Growth,
  J. Math. Sociol. 23  (1999) 255--292. 
%  
\bibitem{hayw05}  J. Hayward,    A General Model of Church Growth and Decline, 
   J. Math. Sociol. 29  (2005) 177--207.

\bibitem{LoganDye84} P.F. Logan, T.W. Dye,  Physics for anthropologists. Search, 15(1984) 30-32.

  \bibitem{roach} P. Ormerod,  A.P. Roach, The medieval inquisition: scale-free networks and the suppression of heresy,  Physica A  339 (2004) 645-652.
  
  \bibitem{Hamilton01} M. Hamilton,  The Sociology of Religion. Theoretical and comparative perspectives, 2nd Ed. %1995
(Routledge, London, 2001) pp. 229-271.
  
 \bibitem{ianna98} L. Iannaccone, Introduction to economics of religion, J. Econ. Lit. 6 (1998) 1465-1496.   

\bibitem{MontrollBadger74}   E. W.  Montroll,  W. W. Badger,  Introduction  to Quantitative Aspects of Social Phenomena   (Gordon and Breach, New York, 1974). 
%
  \bibitem{BainbridgeStark1981} W.S. Bainbridge, R. Stark, The 'consciousness  reformation' reconsidered, J. Sci. Stud. Relig.  20 (1981) 1-16. 
%Social scientists often assume that individuals and societies orient themselves with respect to overarching symbolic frames of reference, sometimes called" meaning systems." But until the recent publication of Robert Wuthnow's analysis of San Francisco area survey data, ... 


 \bibitem{futures}   T.M.  Johnson,  D. Barrett, Quantifying alternate futures of religion  and religions, Futures 36 (2004) 947-960.
 
  


\end{thebibliography}
\end{document}